# Design of Near Infrared and Visible Kinetic Inductance Detectors Using MIM Capacitors

S. Beldi[1] · F. Boussaha[1] · C. Chaumont[1] · S. Mignot[1] · F. Reix[1] · A. Tartari[2] · T. Vacelet[3] · A. Traini[2] · M. Piat[2] · P. Bonifacio[1]

[1] *Laboratoire GEPI, Observatoire de Paris, CNRS, PSL Université, 75014 Paris, France*

[2] *APC, Université de Paris Diderot, 75013 Paris, France*

[3] *LERMA, Observatoire de Paris, CNRS, PSL Université, 75014 Paris, France*

**Abstract**

We are developing superconducting Microwave Kinetic Inductance Detectors to operate at near infrared and optical wavelengths for astronomy. In order to efficiently meet with the requirements of astronomical applications, we propose to replace the interdigitated capacitor by a metal–insulator–metal capacitor which has the advantage of presenting a larger capacitance value within a much smaller space. The pixel will occupy a space of typically $100 \times 85$ μm which is nine times less than a typical pixel size using the interdigitated capacitor operating at the same frequency, below 2 GHz.

## 1 Introduction

During the past 10 years, several groups have been developing MKIDs in order to demonstrate their capability to achieve high-performance detectors from mm, submm/THz through to X-rays. MKID technology allows a straightforward frequency domain multiplexing of a high number of pixels using a single readout line and a basic cryogenic amplification. Therefore, MKIDs technology is likely the best candidate to build cameras comprising thousands of pixels. This is a pivotal feature to achieve high sensitivity, high spatial resolution and high mapping speed for many astronomical applications. To operate at near IR and visible wavelengths, the widely used MKID design, called Lumped Element Kinetic Detector (LEKID), features basically an interdigitated capacitor shorted by a meandered inductive line. In LEKID, the meander features a unique geometry and is optimized to act as an absorber with a high optical efficiency. The interdigitated capacitor is sized to tune the resonance frequency of LEKIDs within the frequency band of the electronic readout system which in our case is between 1 and 2 GHz.

✉ S. Beldi
Samir.Beldi@obspm.fr

The interdigitated capacitor can cover up to 90% of the overall pixel size. This is detrimental for astrophysical applications, for two reasons: only 10% of the focal plane surface occupied by the detector is active and the filling factor is very low. Both facts imply a very low observing efficiency. In order to restore the spatial resolution afforded by the instrument, one has to use dithering. By dithering even a low filling factor detector, that under samples the point spread function, may be employed to produce high-resolution images [1] at the cost of extra observing time. In order to reach a higher filling factor and therefore a larger pixel count in a single array, the current KID size should be diminished typically from hundreds to a few tens of µm and the active fraction of the pixel should be increased. For example, the optical Lumped Element KID developed by [2] has a size $\approx 130 \times 130$ µm for an operating frequency around 4.7 GHz. This pixel can easily reach a size of $300 \times 300$ µm for an operating frequency within 1–2 GHz frequency band. In order to set a lower pixel size, we use instead the parallel plate capacitor with a high dielectric constant [3]. This has another advantage: according to the theoretical model developed by [4, 5], it is possible to increase the detector sensitivity by lowering the excess noise usually observed in MKIDs thanks to the high electric field which can be set inside the parallel plate capacitor through the power driving the resonators.

## 2 Interdigitated Capacitor-Based LEKID

Using the parameters summarized in Table 1, we first simulated an interdigitated capacitor-based LEKID which is illustrated in Fig. 1a. The TiN resonator, defined on a high-resistivity Si substrate, is made up of a meander in parallel of an interdigitated capacitor featuring sizes of $40 \times 40$ µm and $260 \times 300$ µm, respectively. In order to improve the capacitive coupling between the resonator and the 50 Ω niobium (Nb) coplanar waveguide (CPW) readout line, we added an isolated Nb strip line, on the capacitor side, linked to the CPW conductor through an Nb bridge. Figure 1b shows the transmission $S_{12}$ response of the readout line indicating that, as expected, the LEKID resonates at 1.82 GHz.

The interdigitated capacitance value can be approximately estimated by means of SONNET [6] simulations using the approach described elsewhere [7, 8]. Despite its large size, this interdigitated capacitor presents a low value of 1.75 pF.

| Thickness (nm) | 60 |
|:---:|:---:|
| Critical temperature (K) | 1 |
| Kinetic Inductance (pH/□) | 24 |
| Normal resistivity (Ωµ.cm) | 110 |

**Table. 1** Physical and electrical parameters used to simulate our LEKIDs.

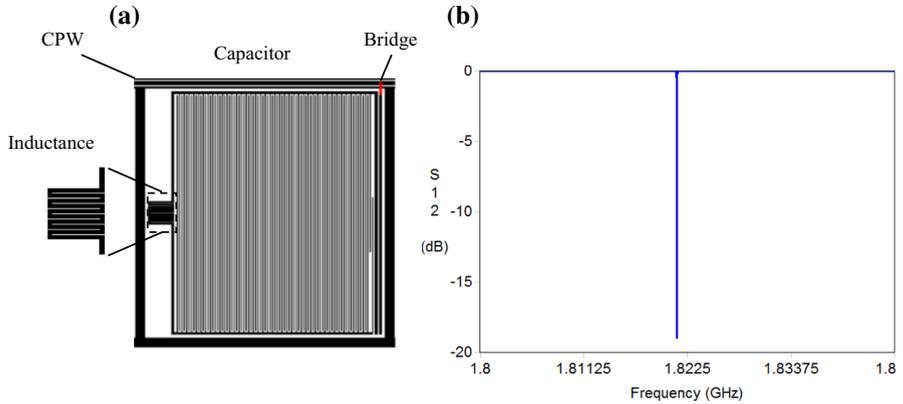

**Fig. 1 a** Sketch of the interdigitated capacitor-based LEKID. **b** Simulated frequency response

| Size of the capacitor (μm) | Capacitance (pF) | | |
|---|---|---|---|
| | Interdigitated capacitor | MIM capacitor ($t_{diel}$=100 nm) | |
| | | Si ($\varepsilon_r$=11) | 9.96) |
| 100 × 40 | 0.08 | 3.89 | 3.04 |
| 100 × 130 | 0.19 | 12.65 | 9.89 |
| 260 × 300 | 1.75 | 75.9 | 59.36 |

**Table 2** Comparison between interdigitated capacitance and MIM capacitance values for three different sizes. The MIM capacitance is calculated using C(pF) 8.85$\varepsilon_r$S/t, where S is the capacitor area, t is the dielectric thickness and $\varepsilon_r$ is the relative permittivity

## 3 MIM Capacitor-Based LEKID

The MIM capacitor allows to achieve higher capacitance thanks to use of high-constant dielectric materials. As an example, Table 2 presents the calculated MIM capacitance values with the available Si and AlN (aluminum nitride) dielectrics compared to the interdigitated capacitance for three different sizes.

Using the same TiN meander, we simulated a LEKID with a MIM capacitor of 40 × 185 μm as shown in Fig. 2. The AlN ($\varepsilon_r$ ~ 8.6) thickness is set to 100-nm-thick allowing resonances to occur at less than 2 GHz.

Changing the value of the MIM capacitance, and therefore the frequency resonance, from one resonator to another, is achieved by removing small 4 × 4 μm squares from the upper electrode of 100 × 40 μm, resulting in spacing between each resonance of 3 MHz as shown in Fig. 3. We can easily multiplex a bit more than 160 pixels without risk of frequency overlap and loss of resonators. Similar results have been obtained with Si ($\varepsilon_r$ ~ 11) and $Al_2O_3$ ($\varepsilon_r$ ~ 9.9). The simulations were carried out with dielectric losses set to 0 which is obviously not true with the most common dielectric films, especially the amorphous ones.

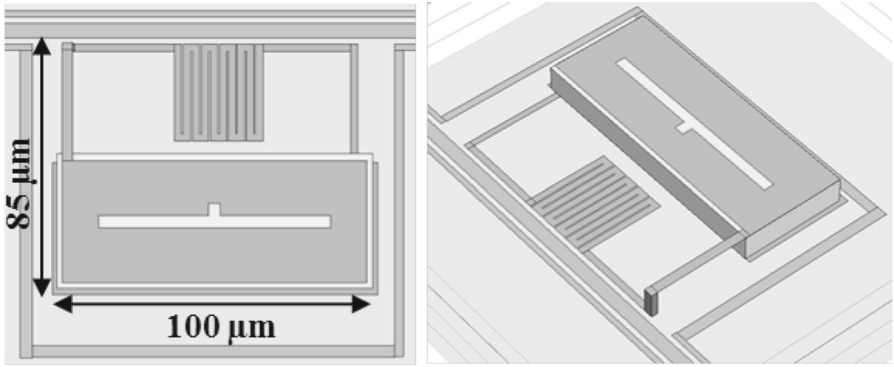

**Fig. 2** MIM-based LEKID design

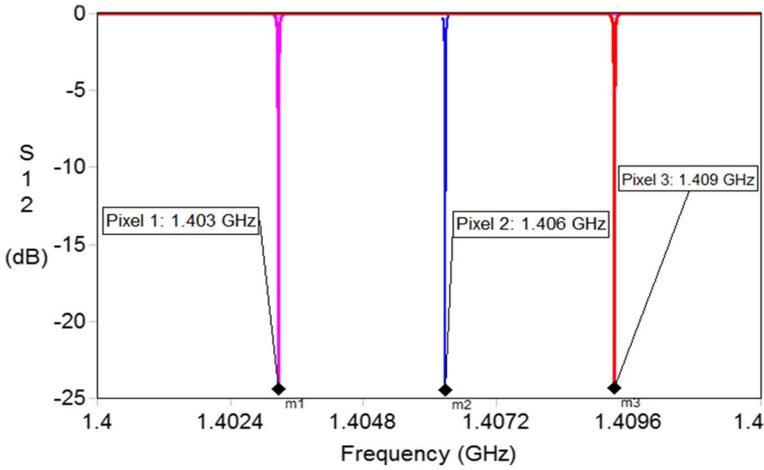

**Fig. 3** SONNET simulations of 3 MIM-based LEKIDs with $\varepsilon_r = 8.6$ and $t = 100$ nm. The resonators have a same inductance (i.e., same meander). The frequency is tuned by changing the upper electrode area of the MIM capacitor

In order to define the dielectric loss threshold to be used for our LEKIDs, we simulated again the above MIM-based LEKID design for different tan δ values. Figure 4 shows the transmission $S_{12}$ response for dielectric losses ranging between $10^{-4}$ and $10^{-6}$. As expected, the resonance quality is deteriorated as dielectric losses increase, featuring a weaker quality factor ($Q_i = 1/\tan\delta$) and a lower resonance depth. Beyond tan $\delta = 10^{-4}$, the resonance is no longer observed. Thus, dielectric losses must be $\leq 10^{-5}$ to achieve a good quality factor ($Q_i \geq 10^5$) and deep resonances. The monocrystalline silicon dielectric which can present a tan δ loss around $5 \times 10^{-6}$ can be one of the best candidates to build high-performance MIM capacitor-based LEKIDs. However, as it is hard to grow monocrystalline silicon layers directly on metallic electrodes, the capacitor can be realized by processing a silicon-on-insulator (SOI) substrate using deep reactive ion etching (DRIE) process as reported in [9]. However, adjusting this process to achieve a large number of LEKIDs is quite challenging. One of the most difficult steps is to electrically connect the bottom electrodes to inductive meanders without affecting the quality and the efficient operation of the LEKIDs. This could be achieved through vertical electrical connections (via) patterned in the silicon layer.

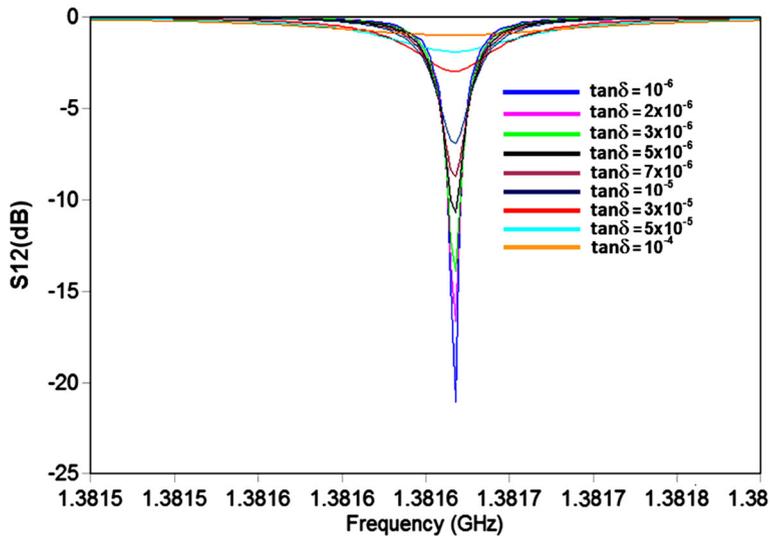

**Fig. 4** SONNET simulation of a MIM-based LEKID for different tan$\delta$

## 4 Conclusion

MIM capacitor-based LEKIDs would considerably reduce the pixel size by a ratio up to 10 [5]. This allows to improve the filling factor. The dielectric losses which would make this approach unsuitable can be likely overcome by, for example, the use of monocrystalline silicon.